\documentclass[12pt]{article}
\usepackage{amsmath,amssymb, amsfonts, amsthm,amscd}
\usepackage[T2A]{fontenc}
\usepackage[cp1251]{inputenc}
\usepackage[russian,ukrainian,english]{babel}
\usepackage{geometry}

\newcommand{\case}[1]{\textbf{Case $\mathbf{#1}$.}\ }

\newtheorem{Theorem}{Theorem}
\newtheorem{Corollary}{Corollary}
\newtheorem{Remark}{Remark}
\newtheorem{Lemma}{Lemma}

\newenvironment{LemmaProof}{\textbf{Proof. }}{\par\noindent\textbf{The Lemma is proved.}}
\newenvironment{TheoremProof}{\textbf{Proof. }}{\par\noindent\textbf{The Theorem is proved.}}

\title{{\Large \textbf{On the parameter $\mu_{21}$ of a complete bipartite graph}}}

\author{\normalsize A.M. Khachatryan$^1$, R.R. Kamalian$^2$}

\date{}

\begin{document}

\maketitle

$\\^1$Ijevan Branch of Yerevan State University, e-mail:
khachatryanarpine@gmail.com $\\^2$The Institute for Informatics and
Automation Problems of NAS RA, \\e-mail: rrkamalian@yahoo.com
\bigskip \bigskip

\begin{abstract}
A proper edge $t$-coloring of a graph $G$ is a coloring of edges of
$G$ with colors $1,2,...,t$ such that all colors are used, and no
two adjacent edges receive the same color. The set of colors of
edges incident with a vertex $x$ is called a spectrum of $x$. An
arbitrary nonempty subset of consecutive integers is called an
interval.

Suppose that all edges of a graph $G$ are colored in the game of
Alice and Bob with asymmetric distribution of roles. Alice
determines the number $t$ of colors in the future proper edge
coloring of $G$ and aspires to minimize the number of vertices with
an interval spectrum in it. Bob colors edges of $G$ with $t$ colors
and aspires to maximize that number. $\mu_{21}(G)$ is equal to the
number of vertices of $G$ with an interval spectrum at the finish of
the game on the supposition that both players choose their best
strategies.

In this paper, for arbitrary positive integers $m$ and $n$, the
exact value of the parameter $\mu_{21}(K_{m,n})$ is found.

\end{abstract}

We consider finite, undirected, connected graphs without loops and
multiple edges containing at least one edge. For any graph $G$, we
denote by $V(G)$ and $E(G)$ the sets of vertices and edges of $G$,
respectively. For any $x\in V(G)$, $d_G(x)$ denotes the degree of
the vertex $x$ in $G$. For a graph $G$, $\delta(G)$ and $\Delta(G)$
denote the minimum and maximum degrees of vertices in $G$,
respectively. For a graph $G$, and for any $V_0\subseteq V(G)$, we
denote by $G[V_0]$ the subgraph of the graph $G$ induced
\cite{West1} by the subset $V_0$ of its vertices. We denote by $C_4$
a simple cycle with four vertices.

An arbitrary nonempty subset of consecutive integers is called an
interval. An interval with the minimum element $p$ and the maximum
element $q$ is denoted by $[p,q]$.

A function $\varphi:E(G)\rightarrow [1,t]$ is called a proper edge
$t$-coloring of a graph $G$, if all colors are used, and for any
adjacent edges $e_1\in E(G)$, $e_2\in E(G)$,  $\varphi(e_1)\neq
\varphi(e_2)$.

The minimum value of $t$ for which there exists a proper edge
$t$-coloring of a graph $G$ is denoted by $\chi'(G)$ \cite{Vizing2}.

For any graph $G$, and for any $t\in[\chi'(G),|E(G)|]$, we denote by
$\alpha(G,t)$ the set of all proper edge $t$-colorings of $G$.

Let us also define a set $\alpha(G)$ of all proper edge colorings of
a graph $G$:
$$
\alpha(G)\equiv\bigcup_{t=\chi'(G)}^{|E(G)|}\alpha(G,t).
$$

If $\varphi\in\alpha(G)$ and $x\in V(G)$, then the set
$\{\varphi(e)/ e\in E(G), e \textrm{ is incident with } x$\} is
called a spectrum of the vertex $x$ of the graph $G$ at the coloring
$\varphi$ and is denoted by $S_G(x,\varphi)$; if $S_G(x,\varphi)$ is
an interval, we say that $\varphi$ is interval in $x$. If $G$ is a
graph, $\varphi\in\alpha(G)$, $R\subseteq V(G)$, then we say, that
$\varphi$ is interval on $R$ iff for $\forall x\in R$, $\varphi$ is
interval in $x$. We say, that a subset $R$ of vertices of a graph
$G$ has an $i$-property iff there exists $\varphi\in\alpha(G)$
interval on $R$. If $G$ is a graph, and a subset $R$ of its vertices
has an $i$-property, we denote by $w_R(G)$ and $W_R(G)$ (omiting the
index in these notations in a peculiar case with $R=V(G)$) the
minimum and the maximum value of $t$, respectively, for which
$\exists\varphi\in\alpha(G,t)$ interval on $R$. If $G$ is a graph,
$\varphi\in\alpha(G)$, then set $V_{int}(G,\varphi)\equiv\{x\in
V(G)/S_G(x,\varphi) \textrm{ is an interval}\}$ and
$f_G(\varphi)\equiv|V_{int}(G,\varphi)|$. A proper edge coloring
$\varphi\in\alpha(G)$ is called an interval edge coloring
\cite{Oranj3, Asratian4, Diss5} of the graph $G$ iff
$f_G(\varphi)=|V(G)|$. The set of all graphs having an interval edge
coloring is denoted by $\mathfrak{N}$.

For a graph $G$, and for any $t\in[\chi'(G),|E(G)|]$, we set
\cite{Mebius6}:
$$
\mu_1(G,t)\equiv\min_{\varphi\in\alpha(G,t)}f_G(\varphi),\qquad
\mu_2(G,t)\equiv\max_{\varphi\in\alpha(G,t)}f_G(\varphi).
$$

For any graph $G$, we set \cite{Mebius6}:
$$
\mu_{11}(G)\equiv\min_{\chi'(G)\leq t\leq|E(G)|}\mu_1(G,t),\qquad
\mu_{12}(G)\equiv\max_{\chi'(G)\leq t\leq|E(G)|}\mu_1(G,t),
$$
$$
\mu_{21}(G)\equiv\min_{\chi'(G)\leq t\leq|E(G)|}\mu_2(G,t),\qquad
\mu_{22}(G)\equiv\max_{\chi'(G)\leq t\leq|E(G)|}\mu_2(G,t).
$$

Clearly, the parameters $\mu_{11}$, $\mu_{12}$, $\mu_{21}$ and
$\mu_{22}$ are correctly defined for an arbitrary graph. Their exact
values are found for simple paths, simple cycles and some
outerplanar graphs \cite{Simple7}, M\"{o}bius ladders
\cite{Mebius6}, complete graphs \cite{Arpine8}, complete bipartite
graphs \cite{Arpine9, Arpine10}, prisms \cite{Arpine11} and
$n$-dimensional cubes \cite{Arpine11, Nikolaev12}. The exact values
of $\mu_{11}$ and $\mu_{22}$ for trees are found in \cite{Evg13}.
The exact value of $\mu_{12}$ for an arbitrary tree is found in
\cite{Trees14}.

In addition to the definitions given above, let us note that exact
values of the parameters $\mu_{12}$ and $\mu_{21}$ have certain game
interpretations. Suppose that all edges of a graph $G$ are colored
in the game of Alice and Bob with asymmetric distribution of roles.
Alice determines the number $t$ of colors in the future coloring
$\varphi$ of the graph $G$, satisfying the condition
$t\in[\chi'(G),|E(G)|]$, Bob colors edges of $G$ with $t$ colors.

When Alice aspires to maximize, Bob aspires to minimize the value of
the function $f_G(\varphi)$, and both of them choose their best
strategies, then at the finish of the game exactly $\mu_{12}(G)$
vertices of $G$ will receive an interval spectrum.

When Alice aspires to minimize, Bob aspires to maximize the value of
the function $f_G(\varphi)$, and both of them choose their best
strategies, then at the finish of the game exactly $\mu_{21}(G)$
vertices of $G$ will receive an interval spectrum.

In this paper, for arbitrary positive integers $m$ and $n$, we
determine the exact value of $\mu_{21}$ for the complete bipartite
graph $K_{m,n}$.

For $m\geq n$, let $K_{m,n}$ be a complete bipartite graph with a
bipartition $(X,Y)$, where $X=\{x_1,\ldots,x_n\}$,
$Y=\{y_1,\ldots,y_m\}$.

Clearly, for any positive integers $m$ and $n$,
$\chi'(K_{m,n})=\Delta(K_{m,n})=m$, $|E(K_{m,n})|=mn$.

First we recall some known results.

\begin{Theorem}\cite{Oranj3, Asratian4, Diss5}\label{Thm1}
If $R$ is the set of all vertices of an arbitrary part of a
bipartite graph $G$, then:
\begin{enumerate}
  \item $R$ has an $i$-property,
  \item $W_R(G)=|E(G)|$,
  \item for any $t\in[w_R(G),W_R(G)]$, there exists $\varphi_t\in\alpha(G,t)$ interval on $R$.
\end{enumerate}
\end{Theorem}

\begin{Theorem}\cite{Vestnik15}\label{Thm2}
For arbitrary positive integers $m$ and $n$,
$w_Y(K_{m,n})=n\cdot\big\lceil\frac{m}{n}\big\rceil$.
\end{Theorem}

\begin{Theorem}\cite{Diss5, Preprint16}
For arbitrary positive integers $m$ and $n$,
$K_{m,n}\in\mathfrak{N}$, $w(K_{m,n})=m+n-gcd(m,n)$,
$W(K_{m,n})=m+n-1$; moreover, for any $t\in[w(K_{m,n}),W(K_{m,n})]$,
there exists $\varphi_t\in\alpha(K_{m,n},t)$ with
$V_{int}(K_{m,n},\varphi_t)=V(K_{m,n})$.
\end{Theorem}

\begin{Corollary}\label{Cor1}
For arbitrary positive integers $m$ and $n$, the inequality
$\max\{m,n\}\leq\min\{m,n\}\cdot\Big\lceil\frac{\max\{m,n\}}{\min\{m,n\}}\Big\rceil\leq
m+n-gcd(m,n)\leq m+n-1\leq mn$ is true.
\end{Corollary}

\begin{Theorem}\cite{Luhansk17}\label{Thm4}
If $G$ is a graph with $\delta(G)\geq 2$,
$\varphi\in\alpha(G,|E(G)|)$, $V_{int}(G,\varphi)\neq\emptyset$,
then $G[V_{int}(G,\varphi)]$ is a forest each connected component of
which is a simple path.
\end{Theorem}

If $G$ is a graph with $\chi'(G)=\Delta(G)$,
$t\in[\Delta(G),|E(G)|]$, $\xi\in\alpha(G,t)$, then for any
$j\in[1,t]$, we denote by $E(G,\xi,j)$ the set of all edges of $G$
colored by the color $j$ at the coloring $\xi$. The coloring $\xi$
is called a harmonic \cite{Arpine9} $t$-coloring of the graph $G$,
if for any $i\in[1,\Delta(G)]$, the set
$$
\bigcup_{1\leq j\leq t,\;j\equiv i(mod(\Delta(G)))}E(G,\xi,j)
$$
is a matching in $G$.

Suppose that $G$ is a graph with $\chi'(G)=\Delta(G)$, $t\in
[\Delta(G),|E(G)|]$, $\xi$ is a harmonic $t$-coloring of $G$. Let us
define a sequence $\xi^*_0,\xi^*_1,\ldots,\xi^*_{t-\chi'(G)}$ of
proper edge colorings of the graph $G$.

Set $\xi^*_0\equiv\xi$.

\case{1} $t=\chi'(G)$. The sequence mentioned above is already
constructed.

\case{2} $t\in[\chi'(G)+1,|E(G)|]$. Suppose that
$j\in[1,t-\chi'(G)]$, and the proper edge colorings
$\xi^*_0,\ldots,\xi^*_{j-1}$ of the graph $G$ are already
constructed. Let us define $\xi^*_j$. For an arbitrary $e\in E(G)$,
set:
$$
\xi^*_j(e)\equiv\left\{
\begin{array}{ll}
\xi^*_{j-1}(e), & \textrm{if $\xi^*_{j-1}(e)\neq \max(\{\xi^*_{j-1}(e)/e\in E(G)\})$}\\
\xi^*_{j-1}(e)-\Delta(G), & \textrm{if $\xi^*_{j-1}(e)=
\max(\{\xi^*_{j-1}(e)/e\in E(G)\})$}.
\end{array}
\right.
$$

\begin{Remark}\label{Rem1}
Suppose that $G$ is a graph with $\chi'(G)=\Delta(G)$,
$t\in[\Delta(G),|E(G)|]$, $\xi$ is a harmonic $t$-coloring of $G$.
All proper edge colorings
$\xi^*_0,\xi^*_1,\ldots,\xi^*_{t-\chi'(G)}$ of the graph $G$ are
determined definitely.
\end{Remark}

\begin{Remark}\label{Rem2}
Suppose that $G$ is a graph with $\chi'(G)=\Delta(G)$,
$t\in[1+\chi'(G),|E(G)|]$, $\xi$ is a harmonic $t$-coloring of $G$.
It is not difficult to see, that for any $j\in[1,t-\chi'(G)]$,
$\xi^*_j$ is a harmonic $(t-j)$-coloring of the graph $G$.
\end{Remark}

\begin{Remark}\label{Rem3}
Suppose that $G$ is a graph with $\chi'(G)=\Delta(G)$,
$t\in[1+\chi'(G),|E(G)|]$, $\xi$ is a harmonic $t$-coloring of $G$.
Assume that $\xi$ is interval in some vertex $z_0\in V(G)$ with
$d_G(z_0)=\Delta(G)$. Then, for any $j\in[1,t-\chi'(G)]$, $\xi^*_j$
is interval in $z_0$.
\end{Remark}

\begin{Lemma}\label{Lem1}
If integers $m$ and $n$ satisfy either conditions $m\geq 3$, $n=2$,
or the inequality $m\geq n\geq 3$, then $\mu_2(K_{m,n},mn)\leq m$.
\end{Lemma}

\begin{LemmaProof}
Assume the contrary. Then there exists
$\varphi_0\in\alpha(K_{m,n},mn)$ with $f_{K_{m,n}}(\varphi_0)=m+k$,
where $k\in[1,n]$. Clearly, $|V_{int}(K_{m,n},\varphi_0)\cap
Y|=m+k-|V_{int}(K_{m,n},\varphi_0)\cap X|$.

\case{1} $|V_{int}(K_{m,n},\varphi_0)\cap X|=0$. In this case we
obtain a contradiction $m<m+k=|V_{int}(K_{m,n},\varphi_0)\cap Y|\leq
|Y|=m$.

\case{2} $|V_{int}(K_{m,n},\varphi_0)\cap X|=1$. In this case
$|V_{int}(K_{m,n},\varphi_0)\cap Y|=m+k-1\geq m\geq 3$. From this
inequality we obtain that
$\Delta(K_{m,n}[V_{int}(K_{m,n},\varphi_0)])\geq 3$, but it is
impossible because of Theorem \ref{Thm4}.

\case{3} $2\leq |V_{int}(K_{m,n},\varphi_0)\cap X|\leq n$. In this
case $|V_{int}(K_{m,n},\varphi_0)\cap
Y|=m+k-|V_{int}(K_{m,n},\varphi_0)\cap X|\geq m-n+k$.

Clearly, if at least one of the inequalities $m-n\geq 1$ and $k\geq
2$ is true, we obtain the inequality
$|V_{int}(K_{m,n},\varphi_0)\cap Y|\geq 2$, which contradicts
Theorem \ref{Thm4}.

Therefore, without loss of generality, we can assume, that $m=n$,
$k=1$, $|V_{int}(K_{m,n},\varphi_0)\cap
Y|=m+1-|V_{int}(K_{m,n},\varphi_0)\cap X|$, $1\leq
|V_{int}(K_{m,n},\varphi_0)\cap Y|\leq m-1$.

Let us notice that the inequality $|V_{int}(K_{m,n},\varphi_0)\cap
Y|\geq 2$ is incompatible with the inequality
$|V_{int}(K_{m,n},\varphi_0)\cap X|\geq 2$ because of Theorem
\ref{Thm4}, therefore $|V_{int}(K_{m,n},\varphi_0)\cap Y|=1$.

An assumption $|V_{int}(K_{m,n},\varphi_0)\cap X|\geq 3$ implies the
inequality $\Delta(K_{m,n}[V_{int}(K_{m,n},\varphi_0)])\geq 3$,
which contradicts Theorem \ref{Thm4}.

An assumption $|V_{int}(K_{m,n},\varphi_0)\cap X|=2$ implies the
equality $f_{K_{m,n}}(\varphi_0)=3$, which is incompatible with the
equality $f_{K_{m,n}}(\varphi_0)=m+k$ because of $m\geq 3$ and
$k=1$.
\end{LemmaProof}

\begin{Lemma}\label{Lem2}
If integers $m$ and $n$ satisfy either conditions $m\geq 3$, $n=2$,
or the inequality $m\geq n\geq 3$, then $\mu_2(K_{m,n},mn)=m$.
\end{Lemma}

\begin{LemmaProof}
It follows from Theorem \ref{Thm1} that there exists
$\widetilde{\varphi}\in\alpha(K_{m,n},mn)$ interval on $Y$. It means
that $\mu_2(K_{m,n},mn)\geq m$. From Lemma \ref{Lem1} we have
$\mu_2(K_{m,n},mn)\leq m$.
\end{LemmaProof}

\begin{Theorem}
For arbitrary positive integers $m$ and $n$, where $m\geq n$,
$$
\mu_{21}(K_{m,n})=\left\{
\begin{array}{ll}
m+1, & \textrm{if $n=1$ or $m=n=2$}\\
m & \textrm{otherwise.}
\end{array}
\right.
$$
\end{Theorem}

\begin{TheoremProof}
\case{1} $n=1$. In this case
$\chi'(K_{m,1})=\Delta(K_{m,1})=|E(K_{m,1})|=m$. It means that for
$\forall\varphi\in\alpha(K_{m,1},m)$, $f_{K_{m,1}}(\varphi)=m+1$.
Hence, $\mu_2(K_{m,1},m)=m+1$, $\mu_{21}(K_{m,1})=m+1$.

\case{2} $m=n=2$. Clearly, $K_{2,2}\cong C_4$, and the theorem
follows from the results of \cite{Simple7}.

\case{3} $m\geq 3$, $n=2$ or $m\geq n\geq 3$.

From Lemma \ref{Lem2} we have $\mu_2(K_{m,n},mn)=m$. Let us show
that for any $t\in[m,mn]$, the inequality $\mu_2(K_{m,n},t)\geq m$
holds.

From Theorems \ref{Thm1} and \ref{Thm2} it follows that for any
$t\in[n\cdot\lceil\frac{m}{n}\rceil,mn]$, there exists
$\varphi_t\in\alpha(K_{m,n},t)$ interval on $Y$ with
$f_{K_{m,n}}(\varphi_t)\geq m$. It means that for any
$t\in[n\cdot\lceil\frac{m}{n}\rceil,mn]$, the inequality
$\mu_2(K_{m,n},t)\geq m$ is true.

Now let us show that for any $t\in[m,m+n-1]$, the inequality
$\mu_2(K_{m,n},t)\geq m$ is also true.

Let us define \cite{Diss5, Preprint16} a proper edge
$(m+n-1)$-coloring $\xi$ of the graph $K_{m,n}$. For any integers
$m$ and $n$, satisfying the inequalities $1\leq i\leq n$, $1\leq
j\leq m$, set $\xi((x_i,y_j))\equiv i+j-1$. It is easy to see that
$\xi$ is a harmonic $(m+n-1)$-coloring of $K_{m,n}$ with
$f_{K_{m,n}}(\xi)=m+n$. Let us consider the sequence
$\xi^*_0,\xi^*_1,\ldots,\xi^*_{n-1}$ of proper edge colorings of
$K_{m,n}$. Taking into account Remarks \ref{Rem1} -- \ref{Rem3}, it
is not difficult to notice, that for any $j\in[1,n-1]$,
$f_{K_{m,n}}(\xi^*_j)=m+n-j$. Consequently, for any $j\in[1,n-1]$,
$f_{K_{m,n}}(\xi^*_j)\geq m+1$.

It means that for any $t\in[m,m+n-1]$, the inequality
$\mu_2(K_{m,n},t)\geq m$ is true indeed.

Now, taking into account Corollary \ref{Cor1}, we can conclude that
for any $t\in[m,mn]$, the inequality $\mu_2(K_{m,n},t)\geq m$ is
proved. From Lemma \ref{Lem2} we obtain $\mu_{21}(K_{m,n})=m$.
\end{TheoremProof}

\begin{Corollary}
For any positive integers $m$ and $n$, where $m\geq n$, the
inequality $m\leq\mu_{21}(K_{m,n})\leq m+1$ holds.
\end{Corollary}

\end{document}